\documentclass[aps,prl,twocolumn,nofootinbib]{revtex4-2}

\usepackage{amsmath}
\usepackage{amssymb}
\usepackage{amsthm}
\usepackage{bm}

\usepackage{xcolor}
\usepackage[colorlinks=true,hyperfootnotes=true,citecolor=cyan]{hyperref}

\providecommand*{\iu}{\ensuremath{\mathrm{i}}}
\def\diff{\mathrm{d}}

\begin{document}

\title{Koopman-von Neumann Field Theory}

\author{James Stokes}
\email{stokesjd@umich.edu}
\affiliation{Department of Mathematics, University of Michigan, Ann Arbor, MI, United States of America}

\date{\today}

\begin{abstract}
The classical many-body problem is reformulated as a bosonic quantum field theory. Quantum field operators evolve unitarily in the Heisenberg picture so that a quantum Vlasov equation is satisfied as an operator identity. The formalism enables the direct transfer of techniques from quantum information and quantum many-body field theory to classical nonequilibrium statistical mechanics. Implications for quantum algorithms are discussed.
\end{abstract}

\maketitle

\textbf{Introduction:}
Shortly after the emergence of quantum theory, Koopman and von Neumann \cite{koopman1931hamiltonian,neumann1932operatorenmethode,neumann1932zusatze} developed a first-quantized Hilbert-space formulation of classical statistical mechanics. The Koopman-von Neumann (KvN) formulation has been a productive source of inspiration, motivating quantum algorithms for simulating classical nonlinear dynamics \cite{joseph2020koopman}, partial differential equations \cite{jin2023time} and quantum-classical dynamics \cite{simon2024improved}, among others. The KvN formalism is also intimately related to functional integral methods for nonequilibrium classical statistical mechanics \cite{gozzi1988hidden,gozzi1989hidden, daus2024field,daus2025cosmic}, collectively known as kinetic field theory (KFT) \cite{bartelmann2016microscopic, bartelmann2017kinetic, fabis2018kinetic, bartelmann2019cosmic, kozlikin2021first, bartelmann2021kinetic, konrad2022asymptotic}.

The nonequilibrium dynamics of an ensemble of indistinguishable particles is a central challenge facing both classical statistical mechanics and quantum mechanics. The difficulty arises from the curse of dimensionality involved in modeling a many-body phase-space density or wave function, respectively. Progress in the quantum many-body problem has been greatly facilitated by the Fock space (quantum field) formalism, which enables the application of powerful analytical and numerical tools, even for number-conserving systems. A comparable quantum field-theoretic formulation of the classical many-body problem would deepen understanding of the quantum-classical connection and help to clarify the potential for quantum advantage in simulating classical nonlinear dynamics. It would also form a practical bridge for importing methods from quantum information science and quantum many-body field theory. This Letter builds that bridge, introducing a KvN field-theoretic formulation of the classical many-body problem in the form of a unitary bosonic quantum field theory characterized by field operators satisfying a quantum Vlasov equation. 

After introducing the theoretical foundations, the remainder of the paper is dedicated to exploring two case studies based on time-dependent perturbation theory and the Dirac-Frenkel variational principle, each showcasing the capability of quantum many-body techniques to inform classical nonequilibrium phenomena.

\textbf{Theory:}
Consider a system of $N$ point particles undergoing Hamiltonian dynamics in Euclidean configuration space, with phase space coordinates denoted by $\bm{x}=(x_1,\ldots,x_N) \in (\mathbb{R}^{2d})^N$ where $x_j = (q_j,p_j) \in \mathbb{R}^{2d}$. 
The statistical mechanics of the system is described by a probability density $\rho(\bm{x},t)$ on phase space, whose time evolution satisfies the Liouville initial value problem
\begin{align}
    \left\{
    \begin{aligned}
    & \frac{\partial\rho}{\partial t}= - \iu L[\rho], &  t & > 0 \\
    & \rho(\bm{x},0) = \rho_0(\bm{x}), & \bm{x} & \in (\mathbb{R}^{2d})^N
    \end{aligned}
    \right.,
\end{align}
where we have defined the partial differential operator,
\begin{align}
    L
    & := \sum_{i=1}^N \left(\frac{\partial H}{\partial p_i} \cdot \frac{1}{\iu} \frac{\partial}{\partial q_i} - \frac{\partial H}{\partial q_i} \cdot \frac{1}{\iu}\frac{\partial}{\partial p_i}\right),
\end{align}
and where $H(\bm{x},t)$ denotes the classical Hamiltonian.
The starting point for the KvN formulation of statistical mechanics is the observation that $L$ is a Hermitian operator on $L^2(\mathbb{R}^{2d})^{\otimes N}$, which satisfies the Leibniz (product) rule. It follows that one can equivalently consider the solution of the Schr\"{o}dinger initial value problem
\begin{align}
    \left\{
    \begin{aligned}
    & \frac{\partial\Psi}{\partial t} = - \iu L[\Psi], &  t & > 0 \\
    & \Psi(\bm{x},0) = \Psi_0(\bm{x}), & \bm{x} & \in (\mathbb{R}^{2d})^N
    \end{aligned}
    \right.,
\end{align}
where the initial condition $\Psi_0 \in L^2(\mathbb{R}^{2d})^{\otimes N}$ satisfies $|\Psi_0|=\sqrt{\rho_0}$.

This Letter is concerned with systems of identical particles, for which the Hamiltonian satisfies the exchangeability assumption
\begin{equation}\label{e:exchangeable}
    \textrm{$H(\bm{x}^\pi,t)=H(\bm{x},t)$ for all $\pi \in S_N$,}
\end{equation}
where $\bm{x}^\pi=(x_{\pi(1)},\ldots,x_{\pi(N)})$. If the initial density $\rho_0$ is also exchangeable, then it follows that the solution of the Liouville equation remains exchangeable for all $t >0$. It is straightforward to show that the exchangeability assumption \eqref{e:exchangeable} implies that
\begin{equation}
    \textrm{$\big[L,U(\pi)\big] = 0$ for all $\pi \in S_N$,}
\end{equation}
where $U(\pi)$ denotes the unitary representation of $\pi \in S_N$ on the Hilbert space $L^2(\mathbb{R}^{2d})^{\otimes N}$. Note that since $L$ commutes with permutations, it can be restricted to symmetric wave functions in $L^2(\mathbb{R}^{2d})^{\otimes N}$.

In order to keep the exposition concrete, we henceforth focus on the specific example of autonomous $N$-body problems with a pairwise parity-invariant interparticle potential,
\begin{equation}
    H = \sum_{i=1}^N \left[\frac{p_i^2}{2m} + U(q_i)\right] + \frac{1}{2}\sum_{i \neq j = 1}^N v(q_i-q_j),
\end{equation}
Then,
\begin{equation}
     L
     =
     \sum_{i=1}^N h(x_i)
     +
     \sum_{i \neq j=1}^N g(x_i,x_j),
\end{equation}
where $h$ and $g$ denote the following Hermitian operators on $L^2(\mathbb{R}^{2d})$ and $L^2(\mathbb{R}^{2d})^{\otimes 2}$, respectively,
\begin{align}
    h(x)
    & := \frac{p}{m} \cdot \frac{1}{\iu}\frac{\partial}{\partial q} - \nabla U(q)\cdot\frac{1}{\iu}\frac{\partial}{\partial p}, \\
    g(x,x')
    & := -\nabla v(q-q') \cdot \frac{1}{\iu}\frac{\partial}{\partial p},
\end{align}
where $x=(q,p)$ and $x'=(q',p')$.
The above considerations make it very natural to identify $L^2(\mathbb{R}^{2d})$ as the single-particle Hilbert space for the many-body bosonic Fock space $\Gamma_+(L^2(\mathbb{R}^{2d}))$. Then $L$ lifts to the following second quantized Hamiltonian $\hat{L}$ acting on the bosonic Fock space,
\begin{equation}
    \hat{L}
    =
    \int_x \hat{\psi}^\dag(x) h(x) \hat{\psi}(x) 
    + 
    \int_{x,x'} \hat{\psi}^\dag(x)\hat{\psi}^\dag(x') g(x,x')\hat{\psi}(x')\hat{\psi}(x),
\end{equation}
where $\hat{\psi}(x)$ is the Schr\"{o}dinger-picture field operator that satisfies
\begin{align}
    \big[\hat{\psi}(x),\hat{\psi}(x')\big]
    & = 0, \\
    \big[\hat{\psi}(x),\hat{\psi}^\dag(x')\big] & = \delta(x-x'), \\
    \hat{\psi}(x)|0\rangle & = 0,
\end{align}
and where $|0\rangle$ denotes the Fock vacuum.
It will be convenient to define the phase-space density operator
\begin{equation}
    \hat{\rho}(x) := \hat{\psi}^\dag(x)\hat{\psi}(x).
\end{equation}
The relation between KvN field theory and the standard first-quantized description can be established by introducing the time-dependent Fock-space state
\begin{equation}
    |\Psi(t)\rangle := \frac{1}{N!}\int \diff\bm{x} \, \Psi(\bm{x},t) \, \hat{\psi}^\dag(x_1) \cdots \hat{\psi}^\dag(x_N) |0\rangle.
\end{equation}
Then the time-dependent Schr\"{o}dinger equation
\begin{equation}\label{e:tdse}
    \frac{\diff}{\diff t}|\Psi(t)\rangle = -\iu \hat{L}|\Psi(t)\rangle,
\end{equation}
is equivalent to the partial differential equation
\begin{equation}
    \frac{\partial \Psi}{\partial t} = -\iu L[\Psi].
\end{equation}

The operator $\hat{L}$ can be interpreted as the Hamiltonian for a $(2d+1)$-dimensional nonrelativistic quantum field theory, expressed in the Schr\"{o}dinger picture. Passing to the Heisenberg picture and recalling that $\hat{L}$ is independent of time, one finds that the field
\begin{equation}
    \hat{\psi}_\textrm{H}(x,t) := \exp(\iu \hat{L} t) \hat{\psi}(x) \exp(-\iu \hat{L} t)
\end{equation}
evolves with time such that the phase-space density operator satisfies the following quantum Vlasov equation \cite{supp},
\begin{equation}\label{e:vlasov}
    \frac{\partial \hat{\rho}_\textrm{H}}{\partial t}(q,p,t) + \frac{p}{m}\cdot \frac{\partial \hat{\rho}_\textrm{H}}{\partial q}(q,p,t) + \hat{F}(q,t) \cdot \frac{\partial \hat{\rho}_\textrm{H}}{\partial p}(q,p,t) = 0,
\end{equation}
where the operator $\hat{F}(q,t)$ is given by
\begin{equation}
    \hat{F}(q,t)
    =
    -\nabla U(q)
    -
    \int\diff q' \left[\int \diff p'\hat{\rho}_\textrm{H}(q',p',t)\right] 
    \nabla v(q-q').
\end{equation}
The formal replacement of the Heisenberg-picture field $\hat{\rho}_\textrm{H}(x,t)$ with a c-number field $\rho(x,t)$ reproduces the classical Vlasov equation. The following section explains the sense in which this substitution is justified.

\textbf{Approximation algorithms:}
Nonequilibrium dynamics is concerned with the time dependence of local operator expectation values, evaluated in some initial state $|\Psi\rangle \in \Gamma_+(L^2(\mathbb{R}^{2d}))$. Consider the simplest case of the phase-space density operator,
\begin{align}\label{e:density}
    \langle \Psi | \hat{\rho}_\textrm{H}(x,t) |\Psi \rangle.
\end{align}
In the following we discuss two approximation algorithms from quantum many-body theory; namely, time-dependent perturbation theory and the Dirac-Frenkel variational principle.

In the absence of an exact solution, a productive strategy is to consider a decomposition of the Hamiltonian into a solvable term, treating the remainder as a perturbation. Identifying $\hat{L}_0$ with the quadratic Hamiltonian and $\hat{L}_1=\hat{L}-\hat{L}_0$, one finds that the Dirac-picture field
\begin{align}
    \hat{\psi}_\textrm{D}(x,t) := \exp(\iu \hat{L}_0 t) \hat{\psi}(x) \exp(-\iu \hat{L}_0 t)
\end{align}
satisfies the equation of motion
\begin{equation}
    \frac{\partial \hat{\psi}_\textrm{D}}{\partial t}(q,p,t) + \frac{p}{m}\cdot \frac{\partial \hat{\psi}_\textrm{D}}{\partial q}(q,p,t) - \nabla U(q) \cdot \frac{\partial \hat{\psi}_\textrm{D}}{\partial p}(q,p,t) = 
    0.
\end{equation}
The above equation of motion can be solved by the method of characteristics. Introduce the flow map $\Phi_t : \mathbb{R}^{2d} \longrightarrow \mathbb{R}^{2d}$, which carries an initial condition $x \in \mathbb{R}^{2d}$ to the solution at time $t \in \mathbb{R}$ of the ordinary differential equation
\begin{equation}
    \dot{x}(t)=V\big(x(t)\big), \quad \quad V(x) := \left(\frac{p}{m},-\nabla U(q)\right).
\end{equation}
It follows that the Dirac and Schr\"{o}dinger pictures are related via
\begin{equation}
    \hat{\psi}_\textrm{D}(x,t) = \hat{\psi}\big(\Phi_{-t}(x)\big).
\end{equation}
Recall that the density operator in the Heisenberg and Dirac pictures is related by
\begin{equation}
    \hat{\rho}_\textrm{H}(x,t) = \hat{U}_\textrm{D}^\dag(t,0) \hat{\rho}_\textrm{D}(x,t) \hat{U}_\textrm{D}(t,0),
\end{equation}
where
\begin{align}
    \hat{U}_\textrm{D}(t,0)
    & = \mathcal{T}\exp \left[-\iu \int_0^t \diff s \, \hat{V}(s)\right],
\end{align}
and where the interaction potential $\hat{V}(s)$ is obtained from $\hat{L}_1$ by replacing $\hat{\psi}(x) \to \hat{\psi}_\textrm{D}(x,s)$.

A convenient choice of initial state is given by a field coherent state, which is a many-body quantum state, parametrized in terms of a single-particle wave function $\phi \in L^2(\mathbb{R}^{2d})$ using a displacement operator as follows
\begin{align}
    |\phi \rangle
    & = D(\hat{\psi},\phi)|0\rangle, \\
    & :=
    \exp\left(\int \diff x \big[ \phi(x) \hat{\psi}^\dag(x) - \phi^\ast(x)\hat{\psi}(x) \big] \right)|0\rangle.
\end{align}
Choosing $\Vert \phi \Vert = 1$ and assuming that $\phi$ is real-valued, so that the initial phase-space density is given by $\varrho(x):=\phi(x)^2$, then a tedious calculation \cite{supp} shows that
\begin{equation}
    \langle \phi | \hat{\rho}_\textrm{H}(x,t) | \phi \rangle
    = \rho_0(x,t) + \rho_1(x,t) + \cdots,
\end{equation}
where
\begin{align}
    \rho_0(x,t) & = \varrho\big(\Phi_{-t}(x)\big), \\
    \rho_1(x,t) & = \int_0^t \diff s \, f\big(\Phi_{s-t}(x),s\big), \\
    f(x,t)
    & =
    \frac{\partial \rho_0}{\partial p}(q,p,t) \cdot \int\diff q' \left[\int \diff p'\rho_0(q',p',t)\right] \nabla v(q-q').
\end{align}
At this order, it can be shown that time-dependent perturbation theory agrees with Vlasov perturbation theory \cite{supp}. Proceeding to higher order captures correlation effects beyond the collisionless limit, analogous to perturbative KFT.

In a first step beyond the perturbative regime, we choose the set of coherent states as an approximation manifold $\mathcal{M}$ for variational quantum dynamics. In particular, consider a parametrized curve $t \to \phi(t) \in L^2(\mathbb{R}^{2d})$ satisfying the differential equation
\begin{equation}
    \frac{\diff}{\diff t}|\phi(t)\rangle = -\iu P_{|\phi(t)\rangle} \hat{L} |\phi(t)\rangle,
\end{equation}
where $|\phi(t)\rangle = D(\hat{\psi},\phi(t))$ and where $P_{|\phi(t)\rangle}$ denotes the projector onto the tangent space $T_{|\phi(t)\rangle}(\mathcal{M})$. Then it can be shown that the function $\rho(x,t):=|\phi(x,t)|^2$ satisfies the classical Vlasov equation,
\begin{equation}
    \frac{\partial \rho}{\partial t}(q,p,t) + \frac{p}{m}\cdot \frac{\partial \rho}{\partial q}(q,p,t) + F(q,t) \cdot \frac{\partial \rho}{\partial p}(q,p,t) = 0,
\end{equation}
where
\begin{equation}
    F(q,t)
    =
    -\nabla U(q)
    -
    \int\diff q' \left[\int \diff p'\rho(q',p',t)\right] 
    \nabla v(q-q').
\end{equation}
The above variational calculation therefore rigorously justifies the c-number replacement in the Heisenberg equation of motion and provides an accompanying a posteriori error estimate \cite[Theorem 1.5]{lubich2008quantum}.

\textbf{Discussion:}
The KvN field theory reformulation of the classical many-body problem opens multiple avenues for theoretical development and practical application.

Extending the formalism to nonautonomous Hamiltonian dynamical systems is a natural next step. Allowing the classical Hamiltonian to depend explicitly on time promotes the Fock space Hamiltonian to a time-dependent operator $\hat{L}(t)$, thereby capturing driven and certain dissipative dynamics such as friction \cite{catli2025exponentially}. Moving beyond the Hamiltonian formalism to general exchangeable dynamical systems can be achieved using a generalized KvN formulation \cite{joseph2020koopman}.

Another natural extension is to expand the approximation manifold $\mathcal{M}$ beyond coherent states. Since the coherent-state manifold $\mathcal{M}_\textrm{Coherent}$ reproduces the collisionless (Vlasov) limit, enlarging the approximation manifold to a superset $\mathcal{M}\supseteq\mathcal{M}_\textrm{Coherent}$ enables the systematic inclusion of collisional physics. Projecting the KvN dynamics onto the Gaussian state manifold, for example, yields coupled evolution equations  analogous to time-dependent Hartree-Fock-Bogoliubov theory. Extending the manifold further with generalized canonical transformations \cite{shi2018variational} would capture higher-order correlation effects.

The second-quantized formulation invites the development of coherent-state functional integral methods, which in turn provide a systematic basis for diagrammatic resummation techniques. Direct comparison with kinetic field theory could clarify the relative strengths of the two approaches.

From a quantum information perspective, KvN field theory suggests intriguing possibilities for characterizing classical ensembles in terms of quantum-information-theoretic tools. It may be instructive to consider mixed-state density operators in the form of noisy classical ensembles, compute reduced density operators for subsystems in phase space, and quantify correlations via entanglement entropy measures.

The KvN field theory offers a new perspective on quantum simulation of classical nonlinear dynamics. Expanding the Schr\"{o}dinger-picture field operator in an orthonormal basis of single-particle wave functions and truncating to finitely many modes reduces the KvN field theory Hamiltonian to a generalized four-local Bose-Hubbard Hamiltonian suitable for quantum simulation algorithms. An important next step is to compare the complexity of simulation with first-quantized KvN \cite{joseph2020koopman, barthe2023continuous}, where the Hamiltonian consists of a polynomial in the quadrature operators of arbitrary degree.

In summary, KvN field theory unifies classical and quantum many-body methods within a single formalism. It provides a conceptual bridge and a practical toolkit for applying quantum information and field-theoretic techniques to problems in classical nonequilibrium statistical mechanics.

\bibliography{references}

\clearpage
\onecolumngrid
\appendix

\section{Supplemental material}

\subsection{Heisenberg equation of motion\label{app:Heisenberg}}
Recall the Heisenberg equation of motion,
\begin{equation}
    \frac{\partial \hat{\psi}_\textrm{H}}{\partial t}(x,t)
    + \iu \big[\hat{\psi}_\textrm{H}(x,t),\hat{H}\big]
    = 0.
\end{equation}
This becomes
\begin{equation}
    \frac{\partial \hat{\psi}_\textrm{H}}{\partial t}(x,t)
    +
    \iu
    h(x) \hat{\psi}_\textrm{H}(x,t)
    +
    \int_{x'}
    \hat{\psi}_\textrm{H}^\dag(x',t)
    \big(
    \iu g(x,x') + \iu g(x',x)
    \big)
    \hat{\psi}_\textrm{H}(x',t)
    \hat{\psi}_\textrm{H}(x,t)
    =
    0,
\end{equation}
where
\begin{align}
    \iu h(x)
    & = \frac{p}{m} \cdot\frac{\partial}{\partial q} - \nabla U(q)\frac{\partial}{\partial p}, \\
    \iu g(x,x')
    & = -\nabla v(q-q') \cdot \frac{\partial}{\partial p}.
\end{align}
Thus
\begin{align}\label{e:psi_eom}
    \frac{\partial \hat{\psi}_\textrm{H}}{\partial t}(q,p,t) + \frac{p}{m}\cdot \frac{\partial \hat{\psi}_\textrm{H}}{\partial q}(q,p,t) + \hat{F}(q,t) \cdot \frac{\partial \hat{\psi}_\textrm{H}}{\partial p}(q,p,t) + \hat{Q}(q,t)\hat{\psi}_\textrm{H}(q,p,t) & = 0,
\end{align}
where
\begin{align}
    \hat{F}(q,t)
    & =
    -\nabla U(q)
    -
    \int\diff q' \left[\int \diff p'\hat{\rho}_\textrm{H}(q',p',t)\right] \nabla v(q-q'), \\
    \hat{Q}(q,t)
    & =
    \int \diff q' \left[\int \diff p'\hat{\psi}_\textrm{H}^\dag(q',p',t)\frac{\partial \hat{\psi}_\textrm{H}}{\partial p'}(q',p',t)\right] \nabla v(q-q').
\end{align}
The operator $\hat{F}$ is manifestly Hermitian whereas $\hat{Q}$ is anti-Hermitian, which follows from integration by parts. Taking the adjoint of \eqref{e:psi_eom} and right-multiplying by $\hat{\psi}_\textrm{H}$,
\begin{align}
    \frac{\partial \hat{\psi}_\textrm{H}^\dag}{\partial t}\hat{\psi}_\textrm{H} + \frac{p}{m}\cdot \frac{\partial \hat{\psi}_\textrm{H}^\dag}{\partial q}\hat{\psi}_\textrm{H} + \frac{\partial \hat{\psi}_\textrm{H}^\dag}{\partial p} \cdot \hat{F}\hat{\psi}_\textrm{H} - \hat{\psi}_\textrm{H}^\dag\hat{Q}\hat{\psi}_\textrm{H} & = 0.
\end{align}
On the other hand, left-multiplying \eqref{e:psi_eom} by $\hat{\psi}_\textrm{H}^\dag$,
\begin{align}
    \hat{\psi}_\textrm{H}^\dag\frac{\partial \hat{\psi}_\textrm{H}}{\partial t} + \frac{p}{m}\cdot \hat{\psi}_\textrm{H}^\dag\frac{\partial \hat{\psi}_\textrm{H}}{\partial q} + \hat{\psi}_\textrm{H}^\dag \hat{F} \cdot \frac{\partial \hat{\psi}_\textrm{H}}{\partial p} + \hat{\psi}_\textrm{H}^\dag\hat{Q}\hat{\psi}_\textrm{H} & = 0.
\end{align}
Adding together gives,
\begin{align}
    \frac{\partial \hat{\rho}_\textrm{H}}{\partial t} + \frac{p}{m}\cdot \frac{\partial \hat{\rho}_\textrm{H}}{\partial q} + \frac{\partial}{\partial p} \cdot \big(\hat{\psi}_\textrm{H}^\dag\hat{F}\hat{\psi}_\textrm{H}\big) & = 0.
\end{align}
Now, assuming Fermi/Bose statistics,
\begin{align}
    \hat{\psi}_\textrm{H}^\dag(x,t) \hat{F}(q,t)
    & = \hat{\psi}_\textrm{H}^\dag(x,t)
    \left[-\nabla U(q)
    -
    \int_{x'}
    \hat{\psi}_\textrm{H}^\dag(x',t)\hat{\psi}_\textrm{H}(x',t) \nabla v(q-q')\right], \\
    & = -\nabla U(q)\hat{\psi}_\textrm{H}^\dag(x,t)
    -
    \int_{x'}
    \hat{\psi}_\textrm{H}^\dag(x',t)
    \hat{\psi}_\textrm{H}^\dag (x,t)
    \hat{\psi}_\textrm{H}(x',t) \nabla v(q-q'), \\
    & = -\nabla U(q)\hat{\psi}_\textrm{H}^\dag(x,t)
    \pm
    \int_{x'}
    \hat{\psi}_\textrm{H}^\dag(x',t)
    \left[
    {\mp}\hat{\psi}_\textrm{H}(x',t)
    \hat{\psi}_\textrm{H}^\dag (x,t)
    \pm\delta(x-x')
    \right]\nabla v(q-q'), \\
    & = 
    \big(\hat{F}(q,t) + \nabla v(0)\big)\hat{\psi}_\textrm{H}^\dag(x,t), \\
    & = 
    \hat{F}(q,t)\hat{\psi}_\textrm{H}^\dag(x,t),
\end{align}
where we used parity invariance of $v(q)$ to set $\nabla v(0)=0$.
\subsection{Time-dependent perturbation theory\label{app:dyson}}
Using a shorthand notation, the commutator becomes
\begin{align}
    \big[\hat{\rho}_\textrm{D}(x,t), \hat{V}(s)\big]
    & =
    \int \diff x_1 \diff x_2
    \big[
    \hat{\rho}_\textrm{D}(x,t),
    \hat{\psi}_\textrm{D}^\dag(x_1,s) \hat{\psi}_\textrm{D}^\dag(x_2,s) g(x_1,x_2) \hat{\psi}_\textrm{D}(x_2,s) \hat{\psi}_\textrm{D}(x_1,s)
    \big], \\
    & = \int_{x_1,x_2} \big[\rho, \psi^\dag_1 \psi^\dag_2 g_{12} \psi_2 \psi_1\big], \\
    & =
    \int_{x_1,x_2} \big(
    \rho \,\psi^\dag_1 \psi^\dag_2 g_{12} \psi_2 \psi_1
    -
    \psi^\dag_1 \psi^\dag_2 g_{12}\psi_2 \psi_1\rho
    \big), \\
    & =
    \int_{x_1,x_2} \big(
    \rho \,\psi^\dag_1 \psi^\dag_2 g_{12} \psi_2 \psi_1
    -
    g_{12}^\ast\psi^\dag_1 \psi^\dag_2 \psi_2 \psi_1\rho
    \big), \\
    & =
    I-I^\dag,
\end{align}
where
\begin{equation}
    I := \int_{x_1,x_2}
    \rho \,\psi^\dag_1 \psi^\dag_2 g_{12} \psi_2 \psi_1.
\end{equation}
Assuming Fermi/Bose statistics,
\begin{align}
    \rho \, \psi_1^\dag \psi^\dag_2
    & =
    \psi_1^\dag \psi^\dag_2 \rho
    +
    \big[\psi,\psi_1^\dag\big]_\pm \psi^\dag \psi_2^\dag 
    \mp
    \big[\psi,\psi_2^\dag\big]_\pm \psi^\dag \psi_1^\dag.
\end{align}
Thus,
\begin{align}
    I
    & = 
    \int_{x_1,x_2}
    \rho \, \psi^\dag_1 \psi^\dag_2 g_{12} \psi_2 \psi_1, \\
    & = \int_{x_1,x_2}
    \big(\psi_1^\dag \psi^\dag_2 \rho
    +
    \big[\psi,\psi_1^\dag\big]_\pm \psi^\dag \psi_2^\dag 
    \mp
    \big[\psi,\psi_2^\dag\big]_\pm \psi^\dag \psi_1^\dag\big) g_{12} \psi_2 \psi_1, \\
    & =
    \psi^\dag \int_{x_1,x_2}
    \big(
    \big[\psi,\psi_1^\dag\big]_\pm \psi_2^\dag 
    \mp
    \big[\psi,\psi_2^\dag\big]_\pm \psi_1^\dag\big) g_{12} \psi_2 \psi_1 +
    \underbrace{\int_{x_1,x_2} \psi^\dag_1 \psi^\dag_2 \, \rho \, g_{12} \psi_2 \psi_1}, \\
    & =
    \psi^\dag \int_{x_1}
    \big[\psi,\psi_1^\dag\big]_\pm
    \int_{x_2} \,
    \psi_2^\dag 
    (g_{12}+g_{21}) \psi_2 \psi_1 +
    \underbrace{\int_{x_1,x_2} \psi^\dag_1 \psi^\dag_2 \, \rho \, g_{12} \psi_2 \psi_1},
\end{align}
where in the last equality we relabeled the dummy integration variables. 
Since the braced term is Hermitian, we obtain (specializing to Bose statistics)
\begin{align}
    & \big[\hat{\rho}_\textrm{D}(x,t), \hat{V}(s)\big] \notag \\
    & =
    \psi^\dag \int_{x_1}
    \big[\psi,\psi_1^\dag\big]
    \int_{x_2} \,
    \psi_2^\dag 
    (g_{12}+g_{21}) \psi_2 \psi_1
    -
    \textrm{h.c.}, \notag \\
    & =
    \hat{\psi}_\textrm{D}^\dag(x,t)
    \int_{x_1}
    \big[\hat{\psi}_\textrm{D}(x,t),\hat{\psi}_\textrm{D}^\dag(x_1,s)\big]  
    \int_{x_2}
    \hat{\psi}_\textrm{D}^\dag(x_2,s) 
    \big(g(x_1,x_2)+g(x_2,x_1)\big) \hat{\psi}_\textrm{D}(x_2,s) \hat{\psi}_\textrm{D}(x_1,s) - \textrm{h.c.} \notag \\
    & = 
    \hat{\psi}_\textrm{D}^\dag(x,t)
    \int_{x_1}
    \big[\hat{\psi}_\textrm{D}(x,t),\hat{\psi}_\textrm{D}^\dag(x_1,s)\big]  
    \int_{x_2}
    \hat{\psi}_\textrm{D}^\dag(x_2,s) 
    \nabla v(q_1-q_2)\cdot\frac{1}{\iu}\left(
    \frac{\partial \hat{\psi}_\textrm{D}}{\partial p_2}(x_2,s) \hat{\psi}_\textrm{D}(x_1,s) - \hat{\psi}_\textrm{D}(x_2,s) \frac{\partial\hat{\psi}_\textrm{D}}{\partial p_1}(x_1,s)\right) - \textrm{h.c.}
\end{align}
Evaluating  the expectation value of the commutator in the coherent state $|\phi \rangle = D(\hat{\psi},\phi)|0\rangle$ we obtain
\begin{align}
    & \langle \phi |\big[\hat{\rho}_\textrm{D}(x,t), \hat{V}(s)\big] | \phi \rangle \notag \\
    & = 
    \phi_0^\ast(x,t)
    \int_{x_1}
    \big[\hat{\psi}_\textrm{D}(x,t),\hat{\psi}_\textrm{D}^\dag(x_1,s)\big]  
    \int_{x_2}
    \phi_0^\ast(x_2,s) 
    \nabla v(q_1-q_2)\cdot\frac{1}{\iu}\left(
    \frac{\partial \phi_0}{\partial p_2}(x_2,s) \phi_0(x_1,s) - \phi_0(x_2,s) \frac{\partial\phi_0}{\partial p_1}(x_1,s) \right) - \textrm{c.c.}
\end{align}
Choosing $\phi$ to be real-valued,
\begin{align}
    \langle \phi |\big[\hat{\rho}_\textrm{D}(x,t), \hat{V}(s)\big] | \phi \rangle
    = 
    2\iu \phi_0(x,t)
    \int_{x_1}
    \big[\hat{\psi}_\textrm{D}(x,t),\hat{\psi}_\textrm{D}^\dag(x_1,s)\big]  
    \int_{x_2}
    \phi_0(x_2,s) 
    \nabla v(q_1-q_2)
    \left(
    \phi_0(x_2,s) \frac{\partial\phi_0}{\partial p_1}(x_1,s)
    -
    \frac{\partial \phi_0}{\partial p_2}(x_2,s) \phi_0(x_1,s) 
    \right).
\end{align}
Now using
\begin{align}
    \phi_0(x_2,s) 
    \frac{\partial \phi_0}{\partial p_2}(x_2,s)
    & =
    \frac{1}{2}
    \frac{\partial \rho_0}{\partial p_2}(x_2,s), \\
    \phi_0(x_1,s) 
    \frac{\partial \phi_0}{\partial p_1}(x_1,s),
    & =
    \frac{1}{2}
    \frac{\partial \rho_0}{\partial p_1}(x_1,s)
\end{align}
we obtain
\begin{align}
    \langle \phi |\big[\hat{\rho}_\textrm{D}(x,t), \hat{V}(s)\big] | \phi \rangle
    & = 
    \iu \phi_0(x,t)
    \int_{x_1} \,
    \big[\hat{\psi}_\textrm{D}(x,t),\hat{\psi}_\textrm{D}^\dag(x_1,s)\big]  
    \int_{x_2}
    \nabla v(q_1-q_2)\cdot
    \left(
    \frac{1}{\phi_0(x_1,s)}\frac{\partial \rho_0}{\partial p_1}(x_1,s)
    \rho_0(x_2,s)
    -
    \phi_0(x_1,s)
    \frac{\partial \rho_0}{\partial p_2}(x_2,s)
    \right), \\
    & = 
    \iu \phi_0(x,t)
    \int_{x_1} \,
    \big[\hat{\psi}_\textrm{D}(x,t),\hat{\psi}_\textrm{D}^\dag(x_1,s)\big]  
    \frac{1}{\phi_0(x_1,s)}\frac{\partial \rho_0}{\partial p_1}(x_1,s)
    \cdot
    \int \diff q_2\left[
    \int \diff p_2 \, \rho_0(q_2,p_2,s)\right]
    \nabla v(q_1-q_2).
\end{align}
Now,
\begin{align}
    \big[\hat{\psi}_\textrm{D}(x,t),\hat{\psi}_\textrm{D}^\dag(x_1,s)\big]  
    & =
    \delta \big(\Phi_{-s}(x_1)-\Phi_{-t}(x))\big), \\
    & 
    = \frac{\delta\big(x_1-\Phi_{s-t}(x)\big)}{|{\det D\Phi_{-s}\big(\Phi_{s-t}(x)\big)}|}, \\
    & = \delta\big(x_1-\Phi_{s-t}(x)\big),
\end{align}
where we have used the fact that the flow map is a volume-preserving diffeomorphism and that it satisfies the group property. Thus
\begin{align}
    \langle \phi |\big[\hat{\rho}_\textrm{D}(x,t), \hat{V}(s)\big] | \phi \rangle
    & = 
    \iu \left.
    \frac{\partial \rho_0}{\partial p_1}(x_1,s)
    \cdot
    \int \diff q_2\left[
    \int \diff p_2 \, \rho_0(q_2,p_2,s)\right]
    \nabla v(q_1-q_2)\right|_{x_1 = \Phi_{s-t}(x)}.
\end{align}
If we define
\begin{equation}
    f(q,p,t)
    =
    \frac{\partial \rho_0}{\partial p}(q,p,t) \cdot \int\diff q' \left[\int \diff p'\rho_0(q',p',t)\right] \nabla v(q-q'),
\end{equation}
then
\begin{equation}
    \langle \phi |\big[\hat{\rho}_\textrm{D}(x,t), \hat{V}(s)\big] | \phi \rangle
    = \iu f\big(\Phi_{s-t}(x),s\big).
\end{equation}
Plugging back into the Dyson expansion gives
\begin{align}
    \langle \phi |\hat{\rho}_\textrm{H}(x,t) | \phi \rangle
    & = 
    \langle \phi | \hat{\rho}_\textrm{D}(x,t) | \phi \rangle
    - 
    \iu \int_0^t \diff s \, \langle \phi |\big[\hat{\rho}_\textrm{D}(x,t), \hat{V}(s)\big] | \phi \rangle
    +
    \cdots, \\
    & = 
    \varrho\big(\Phi_{-t}(x)\big)
    +
    \int_0^t \diff s \,f\big(\Phi_{s-t}(x),s\big)
    +
    \cdots.
\end{align}

Now we compare with Vlasov perturbation theory. Consider the Vlasov equation,
\begin{equation}
    \left\{
    \begin{aligned}
    & \frac{\partial \rho}{\partial t}(q,p,t) + \frac{p}{m} \cdot \frac{\partial \rho}{\partial q}(q,p,t) + F(q,t) \cdot \frac{\partial \rho}{\partial p}(q,p,t) = 0 \\  
    & \rho(q,p,0) = \varrho(q,p)
    \end{aligned}
    \right.
\end{equation}
where
\begin{align}
    F(q,t)
    & =
    -\nabla U(q) -\int\diff q' \left[\int \diff p'\rho(q',p',t)\right] \nabla v(q-q')
\end{align}
Consider a formal expansion in the potential $v$,
\begin{align}
    \rho(x,t) & = \rho_0(x,t) + \rho_1(x,t) + \cdots.
\end{align}
Then $\rho_0$ satisfies
\begin{equation}
    \left\{
    \begin{aligned}
    & \frac{\partial \rho_0}{\partial t}(x,t) + V(x)\cdot\nabla \rho_0(x,t) = 0 \\
    & \rho_0(x,0)=\varrho(x)
    \end{aligned}
    \right.
\end{equation}
which is solved by the method of characteristics giving
\begin{equation}
    \rho_0(x,t) = \varrho\big(\Phi_{-t}(x)\big).
\end{equation}
Now $\rho_1$ satisfies
\begin{equation}
    \left\{
    \begin{aligned}
    & \frac{\partial \rho_1}{\partial t}(x,t) + V(x)\cdot\nabla \rho_1(x,t) = f(x,t) \\
    & \rho_1(x,0)=0
    \end{aligned}
    \right.
\end{equation}
which is solved by
\begin{align}
    \rho_1(x,t) & = \int_0^t \diff s \, f\big(\Phi_{s-t}(x),s\big).
\end{align}
\begin{proof}
\begin{align}
    \frac{\partial \rho_1}{\partial t}(x,t)
    & = 
    f(x,t) +
    \int_0^t \diff s \left( {\frac{\diff}{\diff t} \Phi_{s-t}(x)}\right) \cdot \nabla f\big(\Phi_{s-t}(x),s\big).
\end{align}
Now
\begin{align}
    \frac{\diff}{\diff t} \Phi_{s-t}(x)
    & = - V\big(\Phi_{s-t}(x)\big).
\end{align}
Thus
\begin{align}
    \frac{\partial \rho_1}{\partial t}(x,t)
    & = 
    f(x,t) 
    - \int_0^t \diff s \, V\big(\Phi_{s-t}(x)\big) \cdot \nabla f\big(\Phi_{s-t}(x),s\big).
\end{align}
Now
\begin{align}
    V(x)\cdot\nabla \rho_1(x,t)
    & = \int_0^t \diff s \, \big[D\Phi_{s-t}(x)V(x)\big]^T \nabla f\big(\Phi_{s-t}(x),s\big), \\
    & = \int_0^t \diff s \, V\big(\Phi_{s-t}(x)\big) \cdot \nabla f\big(\Phi_{s-t}(x),s\big),
\end{align}
and thus
\begin{equation}
    \frac{\partial \rho_1}{\partial t}(x,t) + V(x)\cdot\nabla \rho_1(x,t) = f(x,t).
\end{equation}
\end{proof}

\end{document}